\documentclass[twocolumn,showpacs,preprintnumbers,amsmath,amssymb]{revtex4}

\usepackage{graphicx,epsfig}
\usepackage{dcolumn}
\usepackage{bm}

\begin{document}
\topmargin .2cm
\title{DIRECT PHOTON RADIATION FROM A RICH BARYON QUARK-GLUON-PLASMA SYSTEM}
\author{S. S. Singh}
\email{sssingh@physics.du.ac.in.}
\affiliation{ Department of Physics and Astro-Physics, University of Delhi, Delhi - 110007,INDIA}
\begin{abstract}
  The direct photon radiation from a hot 
and interacting fireball system of 
a rich baryon quark-gluon plasma using the Boltzmann 
distribution function for the incoming particles and 
Bose-Einstein distribution for gluon  and
Fermi-Dirac distribution for quark, antiquark and Boltzmann 
distribution for gluon are discussed subsequently.
 The first two distribution functions are chosen for the photon 
production of the first case and the last two functions 
are considered for the second case of my photon production.  
 The thermal photon 
emission rate is 
found that it is infra-red divergent for the massless quarks 
and this divergence in the quark mass is regulated 
using different
cut-off in the quark mass. However , I remove 
this divergence  using the technique of Braaten and Pisarski in the 
thermal mass of  the system in my model of calculation
with a coupling parameter of QGP fireball. 
 Thus, the production rate of the thermal photon is found to be 
smoothly worked by this cut off technique of this model and 
the result is found to be increasing function with 
the increase of the variation in the value of 
chemical potential~  $\mu$.
\end {abstract}
\pacs{ 25.75.Ld, 12.38.Mh, 21.65.+f}
\maketitle
 The study of the fundamental theory of strong 
interactions~\cite{wilczek} is very much important for the present 
scenarios of heavy ion collider physics. This strong 
interaction shows
 a phase transition from normal nuclear matter 
to very high nuclear density
of deconfined phase of quark and gluon 
which are defined as colour exchange 
particles. These deconfined quark and gluon form plasma 
state of matter , the so called Quark-Gluon Plasma (QGP). 
The creation and evolution of such deconfined 
state of colour quark and gluon is due to the central 
collisions of two massive nuclie  and  it has now
become  a subject matter of present day of heavy-ion 
collision at the BNL Relativistic Heavy-Ion collider(RHIC) 
and the CERN Large Hadron collider(LHC). Moreover, the formation 
of this QGP is in time scale of $ ~1 ~fm/c$ after the 
collision~ \cite{Boyanovsky}. 
Perhaps , it is presumed that 
the early universe was in this state upto about a few microsecond after 
the Big-Bang or in the interior core matter of neutron star.
 Today the core program of ongoing relativistic 
nucleus-nucleus collision and heavy ion collision 
is to study the possible 
QGP phase of Quantum Chromodymics(QCD)in which we study the properties 
of the strongly
interacting matter at very high energy density and temperature.
So the experimental measurement and theoretical 
investigation  is one of the basic concepts  in heavy-ion collisions.
Moreover , if plasma is formed in such experiments , then 
there will be questions of emitting  a large number 
of particles
which consist of leptons and photons.  
 These photons and leptons carry  
 high energy~ \cite{shuryak} and these electromagnetic
productions are of much interested as they can not 
interact strongly in the subsequent process of hadronization.
They have
a large mean free path due to the small cross section for electromagnetic
interaction in the plasma . So they carry
the whole informations about the existence of the plasma and it 
is moreover true that over a large range of expected plasma 
temperature , its radiation can be observed throughout its evolution.
On such aspects of electromagnetic radiation , they are considered 
to be the good probes for the formation of QGP.   
\par  
 So far, calculations of the photon production in a QGP 
at finite temperature have been studied and this temperature 
is related to the 
energy density~ $\epsilon$~ given by the stefan boltzmann
~$\epsilon~\sim ~ T^{4}$ and the thermodynamic relation for the system
 is given as : $ T\frac{dp}{dT}- p = \epsilon$~;~ $ p = \frac{1}{3}\sigma T^{4} - A T$ and the linear term is the non-perturbative effect for calculation 
of the pressure and  $'A'$  is a constant parameter.  As indicated ,
most calculations do not involve about the 
chemical potential in electromagnetic production 
even though there is no fully transparency in 
central midrapidity region shown by microscopic models [4,5].
Even at experiments at
AGS and SPS , the chemical potential can not be neglected. 
In this situation ,
the photon production is not only the functions of 
temperature but also the chemical potential. Among the few 
calculations, Dumitru et al.~\cite{Dumitru} calculated the 
photon production in rich baryon density and M.Strickland~ \cite{Strickland}
~too , 
found this thermal photon using J$\ddot{u}$tter 
distribution function . Again , Hammon~ \cite{Hammon} and coworkers 
have indicated that the initial QGP system 
produced at the RHIC energies have finite
baryon density. In this way , Mujumder et al.~\cite{Majumder} and 
Bass et al.~\cite{Bass} have pointed out
the parton rescattering and fragmentation lead to a substantial 
increase in the net-baryon density at midrapidity region.
Very recent work by He Ze-Jun et al.~\cite{He Ze-Jun} is 
one of interesting 
results of thermal photon from this rich baryon QGP too.
Thus, quark chemical potential is very much influencing factor for the 
calculation of this paper. As stated above, the Stefan-Boltzmann function for 
energy  $\epsilon~\sim ~ T^{4}$ , it is modified that the EOS for this case
is expressed as:

\begin{eqnarray}
   (Tlnz)_{f}&=&\left(\frac{g_{f}V}{12}\right) \nonumber\\
       &\times& \left(\frac{7\pi^{2}}{30}T^{4}+\mu^{2}T^{2}+
           \frac{1}{2\pi^{2}}\mu^{4} \right)~, \nonumber\\ 
\\
(Tlnz)_{b}&=&\frac{g_{b}V\pi^{2}}{90}T^{4} ~,
\end{eqnarray}

where these equations give thermodynamic potentials for fermion and boson
particles. $ g_{f} $ and $ g_{b} $ are the respective degeneracies 
that measure the degree of freedom for fermion and boson.
On this thermodynamic potential , the number density and 
energy density are obtained.

\begin{equation}
n = \frac{1}{V}\partial{(T\,lnZ)}/{\partial{\mu}} ~;~ \epsilon=T\partial{(T\, lnZ)}/{\partial{T}}+\mu n
\end{equation}

In this paper , I did the calculation of the photon radiation at 
transition phase around the temperature
$ T=(0.15-0.17) GeV $  as well as the hot phase of QGP system
 i.e the temperature $T=0.25 GeV$ and its comparision about these two
results when I use different distribution functions
for the calculation at these temperatures. 
 
The process of these finding is being done by using QCD annihilation 
and compton process between quark , antiquark and gluon by using 
different distribution functions for quark , antiquark and gluon.
I used the 
mass of
the quark as effective mass with the coupling parameter which is 
obtained through
QGP fireball formation. The coupling parameter is calculated with
this paramerlised factor $\gamma_{q,g}$$[15]$.
\begin{equation}
  \gamma_{q,g}=\sqrt{2}\sqrt{\frac{1}{\gamma_{q}^{2}}
       +\frac{1}{\gamma_{g}^{2}}}~ ,
\end{equation}
where $\gamma_{g}=a \gamma_{q}$ and $'a'$ is either $6$ or $8$ with
$\gamma_{q}=1/6$
\par
{\bf Photon production from a rich baryon QGP}: 
The calculation of photon production from a rich 
baryon QGP is found to be very
interesting theoretical problem. So I begin the calculation 
of virtual photon yield through the Drell-Yan Mechanism
as :
$ q\bar{q} \rightarrow \gamma g $, $ q(\bar{q}) g \rightarrow \gamma q(\bar{q})$ . In this mechanism , the different distribution
 functions are used for the incoming particles i.e quark and antiquark. 
In the first process of the calculation , Boltzmann distribution function 
$f_{q}(E_{q})=exp(-\frac{E\pm\mu}{T})$ is used
for the incoming particle and Bose-Einstein distribution 
$f_{g}(E_{g})=\frac{1}{exp(E/T)-1}$ for gluonic particle.
The choice of such functions is attributed as the usual pattern
of the most calculations .Thus, In my case ~\cite{wong}, the photon yield rate 
per space volume for the 
annihilation process is given below :
\begin{eqnarray}
        E_{\gamma}\frac{dN^{ann}}{dP_{\gamma} d^{4}X} &=& 4 N_{s}^{2}\sum_{f=1}^{N_{                         f}}\frac{1}{(2 \pi)^{6}} 
                                          \int d^{3}p_{1} d^{3}p_{2} E_{\gamma} \nonumber \\
            &\times&f_{1}f_{2}
             (1+ f_{3})
              \left|M_{i}\right|^{2} ~,
\end{eqnarray}
where $ f_{i} $ are the particle distribution functions 
of $ E_{i} , \mu $ and $ T $ , $ i = 1 , 2 , 3 $ denotes 
the quark , antiquark and gluonic
 particles and $\left|M_{i}\right|^{2}$ is 
the amplitute for the corresponding collision which is given as:
 $\left|M_{i}\right|^{2} = \frac{d\sigma}{dp_{\gamma}}(q_{1}\bar{q}_{2}\rightarrow \gamma g) v_{q \bar{q}}$
and $ N_{s} $ is degree of freedom of the quark.
By pluging the
corresponding distribution function in the above integration ,
it obtains the photon yields through annihilation process.
\begin{eqnarray}
E_{\gamma}\frac{dN^{ann}}{dP_{\gamma} d^{4}X}=\frac{5 \alpha_{e}\alpha_{s}}{27\pi^{2}}
T^{2}\exp(-\frac{E}{T}) \nonumber \\
 \times\lbrace \ln(\frac{4 E T}{m_{q}^{2}})-C_{F}-1 
 -\frac{6}{\pi^{2}}\sum_{n=1}^{\infty}\frac{\ln(n)}{n^{2}} 
\rbrace
\end{eqnarray}

and this expression is independent of the chemical potential.                                                                                
But in the compton process, it can be performed through these two reactions :
$ q g\rightarrow \gamma q $ ;  $ \bar{q} g\rightarrow \gamma \bar{q}$.

In a similar way,
it  does the compton process with the same frequency distribution and its
photon yields as:
\begin{eqnarray}
        E_{\gamma} \frac{dN^{comp}}{dP_{\gamma}d^{4}X} &=& 4 N_{s}N_{\epsilon}\sum_{f=1}^{N_{                         f}}\frac{1}{(2 \pi)^{6}}
                                          \int d^{3}p_{1}
   d^{3}p_{2}E_{\gamma} \nonumber \\
            &\times&f_{1}f_{3}
             (1- f_{2})
              \left|M_{i}\right|^{2}
\end{eqnarray}
Here , again
$\left|M_{i}\right|^{2}$ is
the amplitute for the corresponding collision which is given as:
 $\left|M_{i}\right|^{2} = \frac{d\sigma}{dp_{\gamma}}(g q(\bar{q}_{2})\rightarrow \gamma q(\bar{q}) v_{gq (\bar{q})}$ ,

 where, 
$v_{gq (\bar{q})}$ is the relative velocity between the quark and gluon.
$ N_{s}$ and $ N_{\epsilon} $, in this case too, indicate the degree of 
freedom for their interactions. Thus , by pluging the corresponding
parameter the photon production or yield per
space volume is given :
\begin{eqnarray}
E\frac{dN^{comp}}{dP_{\gamma} d^{4}X}=\frac{5 \alpha_{e}\alpha_{s}}{9\pi^{4}}
T^{2}\exp(-\frac{E}{T})[(1\mp\exp(\pm\frac{\mu}{T}))\frac{6}{\pi^{2}}  \nonumber \\\times \lbrace \ln(\frac{4 E T}{m_{q}^{2}})-C_{Euler}-\frac{1}{2} \rbrace \nonumber \\
 -\frac{6}{\pi^{2}}(1\mp\exp(\pm\frac{\mu}{T}))\sum_{n=1}^{\infty}\frac{\ln(n)}{n^{2}} \nonumber \\ +\exp(\pm\frac{\mu}{T}) \lbrace \ln(\frac{4 E T}{m_{q}^{2}})-C_{Euler}-\frac{1}{2}
\rbrace] ~,
\end{eqnarray}
  where $C_{Euler}$ is Euler number  $0.577215 $ . 
In the above equation , the first polarity $\mp $ shown in above
equation , the upper one is compton process for 
the quark and lower one is for antiquark and the second polarity
    $\mp $ shown second in the equation
is for antiquark and quark . So now the final production 
rate for  both cases of
annihilation as well as compton process is shown in Fig.$ 1 $  and  $2 $ 
subsequently. This calculation of photon production through
annihilation for the hot QGP 
system as well as transition temperature is shown in the Fig. $1$.

Again for the second calculation of photon yield, I proceed 
the same technique through annihilation and compton process by using 
the Fermi-Dirac distribution functions 
$f_{i}(E_{i})=\frac{1}{exp(\frac{E_{i}\pm\mu}{T})-1}$ ~, 
here $'i'$ denotes
for the quark and antiquark for strong interaction
and Boltzmann distribution function indicated above is used for gluon. 
I like 
to compare this results with the previous results as mentioned in 
the first case.
Pluging the corresponding frequency distribution in eqn.(5) and (7) for 
annihilation and compton process , I obtain the photon production rate
per space volume and found to be as : 
\begin{eqnarray}
E\frac{dN^{ann}}{dP_{\gamma} d^{4}X}=\frac{5 \alpha_{e}\alpha_{s}}{9\pi^{4}} 
T^{2}[\frac{1}{\exp(-\frac{E-\mu}{T})+1}  \nonumber \\\times [ \lbrace \ln(\frac{4 E T}{m_{q}^{2}})-C_{Euler}-1 \rbrace \nonumber \\ 
 +(1-\exp(\frac{\mu}{T}))\sum_{n=1}^{\infty}\frac{(-1)^{n+1}}{n^{2}}\exp(-\frac{\mu n}{T}) \nonumber \\ \lbrace \ln(\frac{4 E T}{m_{q}^{2}})-C_{Euler}-1-\ln(n)\rbrace ] \nonumber \\ +\frac{1}{\exp(E+\mu)/T+1}[\lbrace\ln(\frac{4 E T}{m_{q}^{2}})-C_{Euler}-1 \rbrace \nonumber \\ 
+(1-\exp(-\frac{\mu}{T})\sum_{n=1}^{\infty}\frac{(-1)^{n+1}}{n^{2}}\exp(-\frac{\mu n}{T} \nonumber\\ 
\times \lbrace \ln(\frac{4 E T}{m_{q}^{2}})-C_{Euler}-1-\ln(n)\rbrace ] ]  ~,
\end{eqnarray}

where 
  $C_{Euler}$  is the same Euler number and $\alpha_{e}=1/137$.
\\
\\ However , in this annihilation process ,
there are two major parts of the contribution in the photon yield.
One is due to quark and another is from antiquark. This result 
too is shown in Fig.$3$
and the corresponding calculated photon yield rate 
through compton process is :
\begin{eqnarray}
E\frac{dN^{comp}}{dP_{\gamma} d^{4}X}=\frac{10 \alpha_{e}\alpha_{s}}{9\pi^{4}}
T^{2} \frac{\exp(\pm \frac{\mu}{T})}{\exp(\frac{(E \pm \mu)}{T}+1}) \nonumber \\
 \times\sum_{n=1}^{\infty}\frac{(-1)^{n+1}}{n^{2}}\exp(\mp \frac{\mu n}{T}) \nonumber \\ \lbrace \ln(\frac{4 E T}{m_{q}^{2}})-C_{Euler}+1/2-\ln(n)\rbrace
\end{eqnarray}
Again , here the polarity before the exponential function ,
the upper one is for quark and lower one is for antiquark.
The above expressions   show the computed thermal spectrum results. 
There is in need of setting the quark mass in all photon yield
expressions as the massless quark shows divergent in the infrared region.
So the quark mass is replaced by infrared cut-off which is 
$ 2 k_{c}^{2}$. This replacement is being done through the 
technique of Braaten and Pisarski~\cite{Braaten} 
and $ k_{c}^{2} = \frac{1}{6} g^{2} T^{2}$
where $'g'$ is QCD coupling parameter . This parameter is obtained through
QGP fireball formation model. The magnitude is given :

\begin{equation}
    g^{2}=\frac{16\pi}{27}
         \frac{1}
          {
           ln(1+\frac{k^{2}}{\Lambda^{2}})
          }
\end{equation}

 with the QCD parameter $\Lambda=0.15~ GeV$ and
    $ k=(\frac{\gamma_{q,g} N^{\frac{1}{3}}
      T^{2}\Lambda^{2}}{2})^{\frac{1}{4}}$
 is low momentum cut-off value with $N=\frac{16 \pi}{27}$
where $\gamma_{q,g}$ is the phenomological
flow parameter $[15]$ defined above to take care of the hydrodynamical
aspects of the hot as well as transition QGP droplets . 
It is obviously found that g is
approximately equal to $1.29 $ which is slightly strong compared to the
other calculations and $\alpha_{s}=\frac{g^{2}}{4 \pi}$.

\begin{figure}
\resizebox*{3.1in}{3.1in}{\rotatebox{270}{\includegraphics{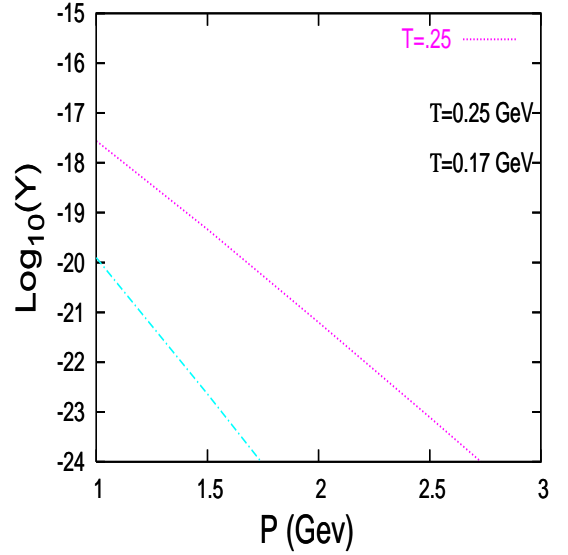}}}
\vspace*{0.5cm}
\caption[]{
The photon emission rate through ann,$Y=E\frac{dN^{ann}}{dPd^{4}X}$, log($Y$)
at
thermal temperature $ T=0.25~ GeV~ and~  0.17~ GeV $ with the parameter
$\gamma_{q,g}$ through the  Boltz.dist.for incoming quark , antiquark 
$\bar{q}$.
and Bose-Ein.dist.for gluon}
\label{scaling}
\end{figure}
\begin{figure}
\resizebox*{3.1in}{3.1in}{\rotatebox{270}{\includegraphics{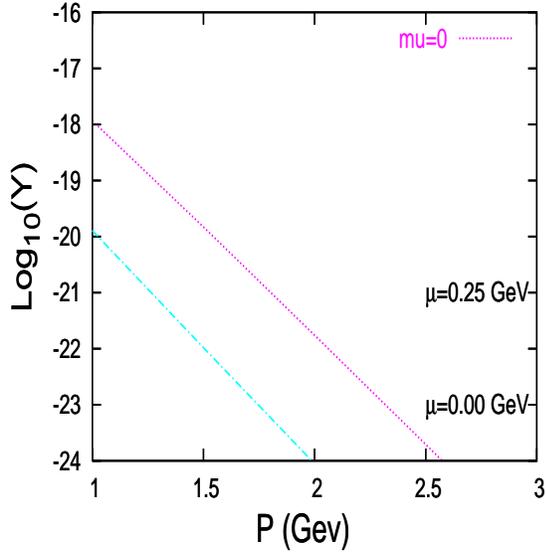}}}
\vspace*{0.5cm}
\caption[]{
The photon emission rate through comp,$Y=E\frac{dN^{comp}}{dPd^{4}X}$,
log($Y$) at thermal temperature
$ T=0.25~ GeV $ with the parameter
$\gamma_{q,g}$ through Boltz dist. for quark $ q $
 and Bose-Ein.dist.for gluon.
}
\label{scaling}
\end{figure}
                                                                                
\begin{figure}[h]
\resizebox*{3.1in}{3.1in}{\rotatebox{270}{\includegraphics{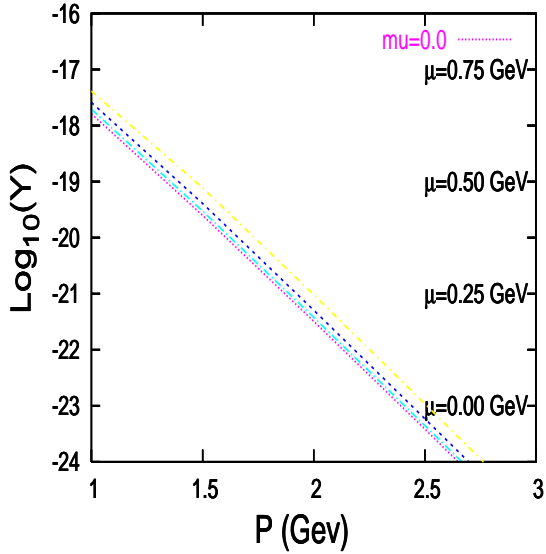}}}
\caption[]{
The photon emission rate through ann,$Y=E\frac{dN^{ann}}{dPd^{4}X}$, log($Y$) at 
thermal temperature $ T=0.25 ~GeV  $  with this parameter 
$\gamma_{q,g}$ through Fermi Dirac for $q$ and $\bar{q}$ and Boltz.dist.
for gluon.
}
\label{scaling}
\end{figure}
\begin{figure}[h]
\resizebox*{3.1in}{3.1in}{\rotatebox{270}{\includegraphics{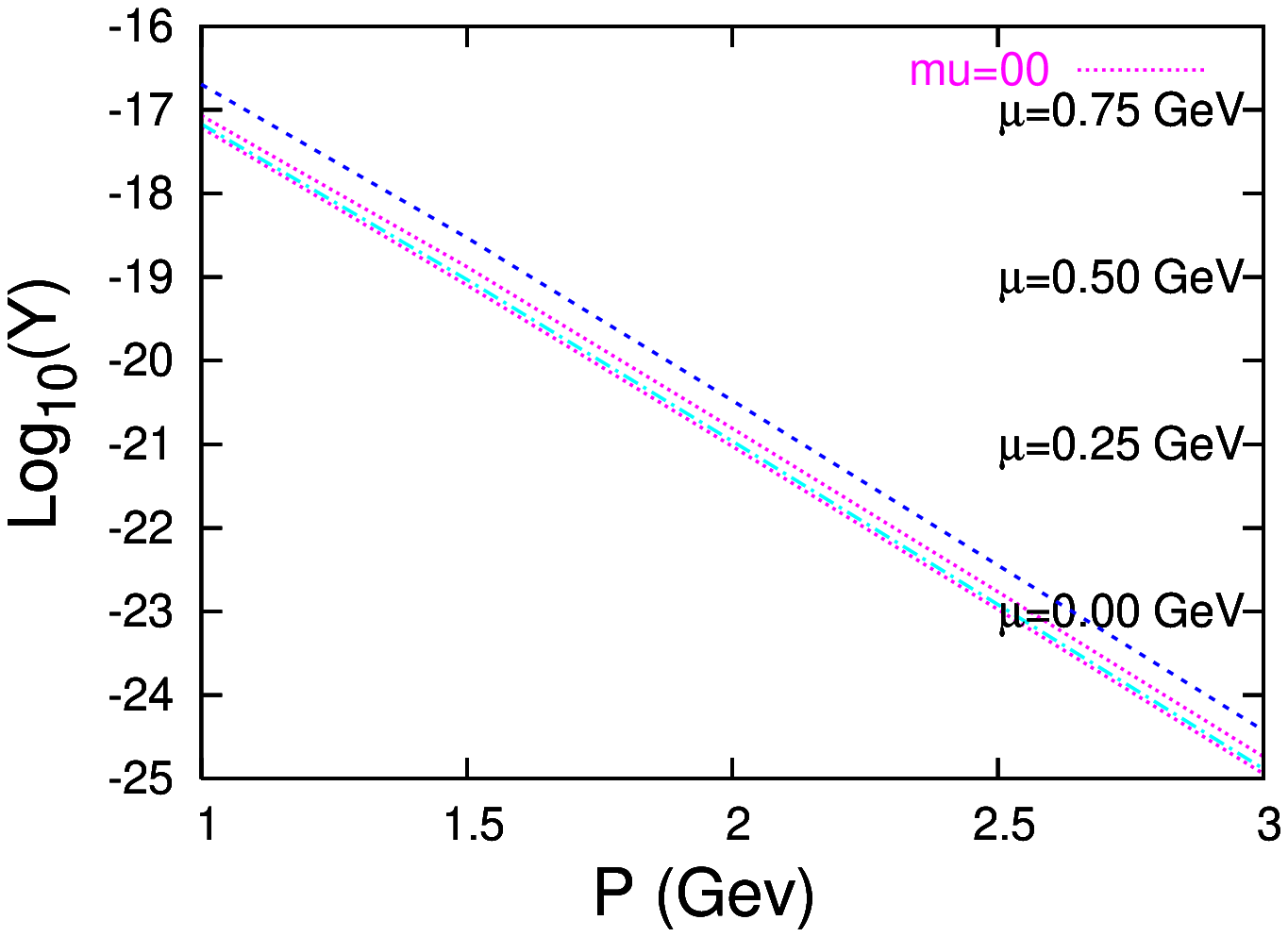}}}
\caption[]{
The photon emission rate through comp, $Y=E\frac{dN^{comp}}{dPd^{4}X}$. 
log($Y$) at 
thermal temperature $ T=0.25 ~GeV $  with this parameter
$\gamma_{q,g}$ through Fermi-dirac for quark $q$ and Boltzmann dist. 
for gluon.
}
\label{scaling}
\end{figure}
\begin{figure}[h]
\resizebox*{3.1in}{2.5in}{\rotatebox{270}{\includegraphics{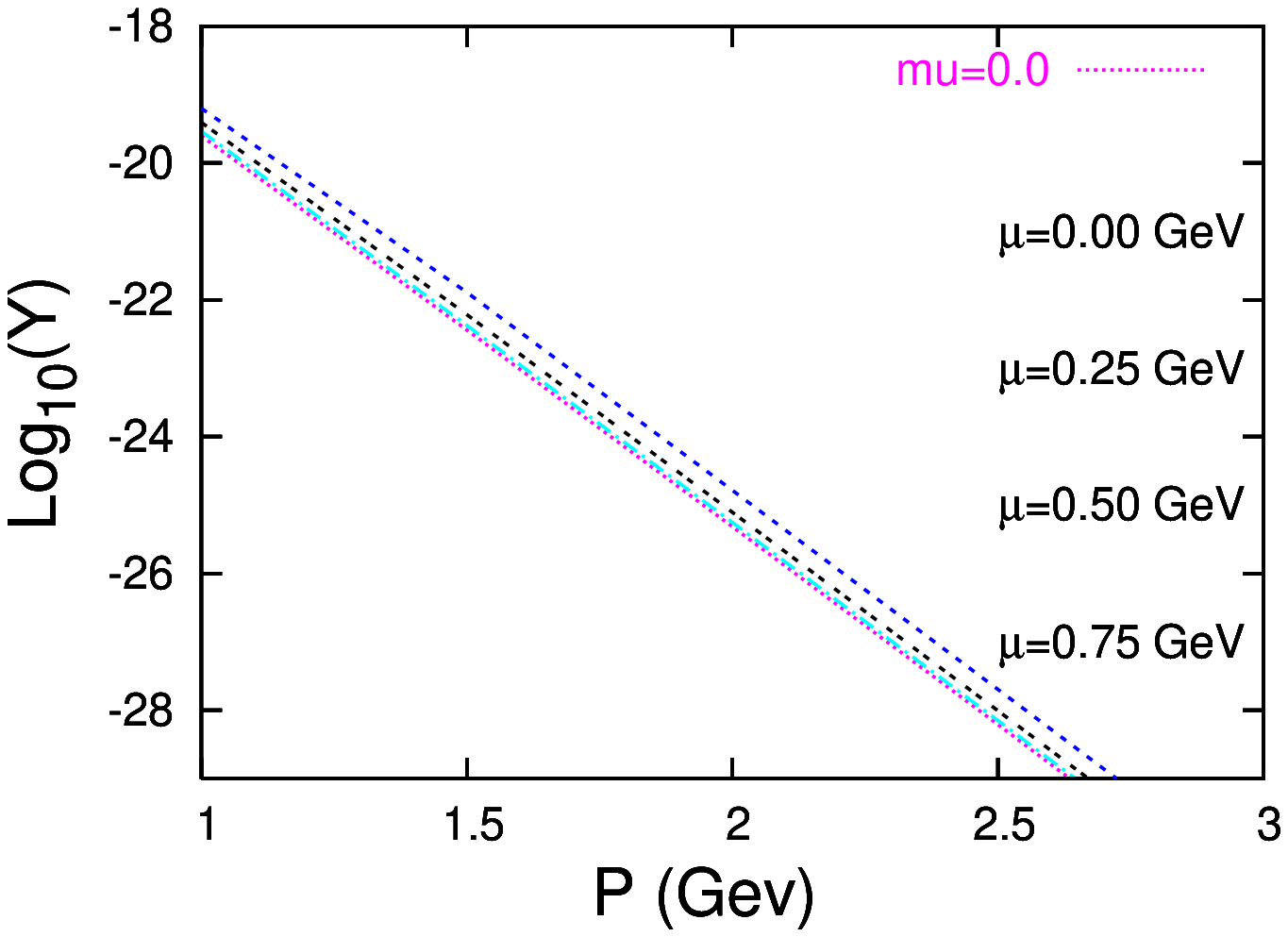}}}
\caption[]{
The photon emission rate through comp,
 $Y=E\frac{dN^{comp}}{dPd^{4}X}$,
log($Y$) at
thermal temperature $ T=0.17 ~GeV $  with this parameter
$\gamma_{q,g}$ through Fermi-dirac for quark $q$ and Boltzmann dist.
for gluon. 
}
\vskip 1.8cm
\label{scaling}
\end{figure}
\begin{figure}[h]
\resizebox*{3in}{2.5in}{\rotatebox{360}{\includegraphics{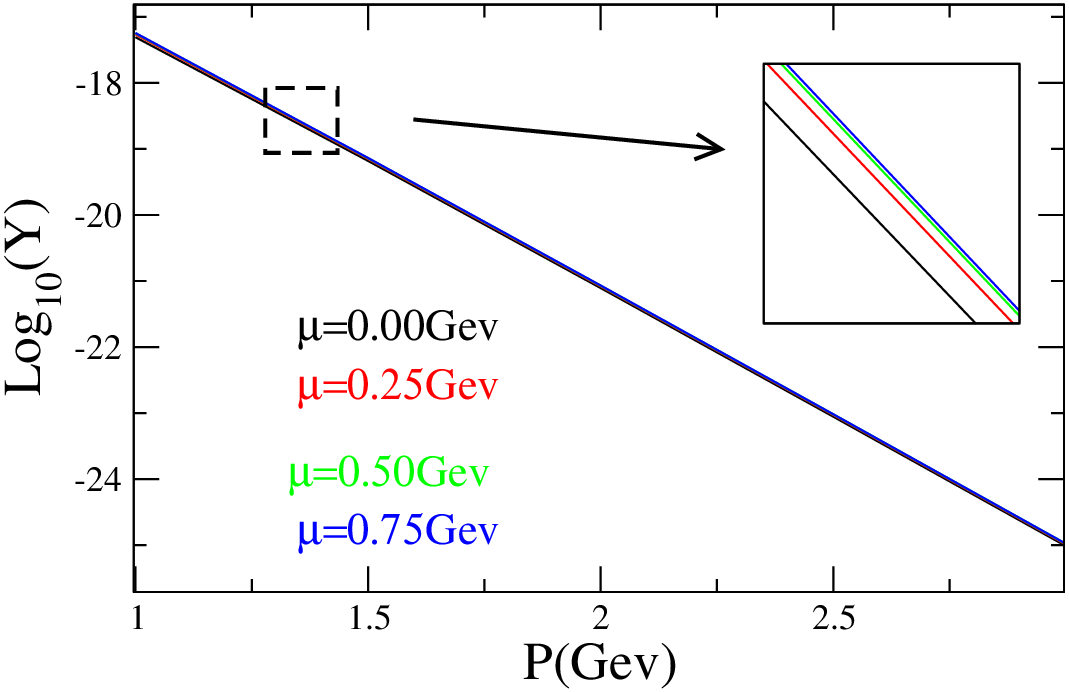}}}
\caption[]{
The photon emission rate through comp, 
$Y=E\frac{dN^{comp}}{dPd^{4}X}$, log($Y$) at 
thermal temperature $ T=0.25 ~GeV $ with the parameter 
$\gamma_{q,g}$ through Fermi-dirac for antiquark $\bar{q}$ and Boltz.dist.for
gluon.
}
\label{scaling}
\end{figure}
\par
{\bf Results and Conclusions}: In this present paper , I attempt to 
evaluate the photon production rate through two processes using 
the Boltzmann distribution function and Bose-Einstein distribution 
function for the incoming particles quark, antiquark and gluon for hot QGP 
system as well as just around transition phase region for different 
coupling parameters based on $\gamma_{g}=6 \gamma_{q}$ or 
$\gamma_{g}=8 \gamma_{q}$ in the first calculation . 
The photon production rate for both of 
these coupling parameters
 are found to be almost similar in both the calculations 
of annihilation as well as the comptom process in this first case. 
 So either this 
$\gamma $ factor with $\gamma_{g}=6 \gamma_{q}$ or
$\gamma_{g}=8 \gamma_{q}$ is very important for photon production rate 
in both
cases with this two temperatures ~\cite{Singh}. In the annihilation 
process of the first case, the 
chemical potential does not play any role for both of the temperatures
but as the temprature rises , the photon production is less suppressed with
the high temperature which is clearly defined in Fig.$1$.
In the case of compton process , there is  some effect in producing 
the photon when the chemical potential is very high compared to
the temperature. There is diverging even though by the effective 
mass of the quark with coupling parameter at low temperature. 
Only when the chemical
potential is around ~$\mu=0.25~GeV$ , the result can be obtained at 
the hot QGP. At transition temperature , we can not see this photon 
radiation.
\par
 In the second case , the photon radiation through annihilation
is found  to be sharp with the chemical potential. With the increase
in the value of~ $\mu $ , the photon radiation is increased  
in the hot QGP shown in Fig. $3$ 
and still divergence is coming out for the transition 
temperature. But the compton process such as $q g\rightarrow \gamma q$ , 
the result is shown in Fig. $4$ for the case of the hot QGP.
Again , it follows the behaviour of the annihilation with less suppression. 
Now ,
the photon suppression for the transition temperature is still 
higher as compared to the hot plasma in compton process of the second case
of distribution function. 
If we use the compton
process for $\bar{q}g\rightarrow \gamma \bar{q}$ , there is no much
distiction about the suppression with the chemical potential
even at high temperature without resolution in the scale. 
Looking into its resolution , 
I can see 
their distiction with effect of the chemical potential $\mu $. 
Moreover , it is necessary to look
into the case of this photon production due to the transverse momentum
$ P $ factor too. Only the transverse momemtum $P$ plays
the role in the case of annihilation process of 
my first calculation. There is a wide difference in their photon
production  with the low value of the transverse momentum. 
It can not observe much difference in the second case of 
my calculation for both annihilation and compton process. Above all , 
photon production or suppression
is observed in both the cases of temperature for
these distribution function in the second case . But it is well contributed
by the hot QGP. It shows that
there are emission of electromagnetic radiation in the hot QGP. 
  
\acknowledgements
  We are very thankful to Dr. R. Ramanathan and Dr.K.K.Gupta 
for their constructive 
suggestions and discussions.

\end{document}